\documentclass[conference]{IEEEtran}
\IEEEoverridecommandlockouts
\usepackage{cite}
\usepackage{amsmath,amssymb,amsfonts}
\usepackage{graphicx}
\usepackage{textcomp}
\usepackage{xcolor}
\usepackage{blindtext}
\usepackage{lettrine}
\usepackage{csvsimple}
\usepackage{caption}
\usepackage{subcaption}
\usepackage{authblk}
\usepackage{mathrsfs} 
\usepackage{float}

\makeatletter
 \let\old@ps@headings\ps@headings
 \let\old@ps@IEEEtitlepagestyle\ps@IEEEtitlepagestyle
 \def\confheader#1{%
 \def\ps@headings{%
 \old@ps@headings%
 \def\@oddhead{\strut\hfill#1\hfill\strut}%
 \def\@evenhead{\strut\hfill#1\hfill\strut}%
 }%
 \def\ps@IEEEtitlepagestyle{%
 \old@ps@IEEEtitlepagestyle%
 \def\@oddhead{\strut\hfill#1\hfill\strut}%
 \def\@evenhead{\strut\hfill#1\hfill\strut}%
 }%
 \ps@headings%
 }
\makeatother

\newcommand\blfootnote[1]{%
  \begingroup
  \renewcommand\thefootnote{}\footnote{#1}%
  \addtocounter{footnote}{-1}%
  \endgroup
}

\confheader{%
\small{\textit{32nd IEEE International Symposium on Space THz Technology (ISSTT 2022)\ \ \ \ \ \ \ \ \ \ \ \ \ \ \ \ \ \ \ \ \ \ \ \ \ \ \ \ \ \ \ \ \ \ \ \ \ \ \ \ Baeza, Spain, October 16-20, 2022}
}}
\begin{document}
\bstctlcite{IEEEexample:BSTcontrol}
\author{ Adrian K. Sinclair, James R. Burgoyne, Yaqiong Li, Cody Duell, Scott C. Chapman, Anthony I. Huber,\\ and Ruixuan Xie
}
\title{Breaking the 10 mW/pixel Limit for Kinetic Inductance Detector Readout Electronics}

\maketitle

\begin{abstract}
We demonstrate a prototype kinetic inductance detector (KID) readout system that uses less than 10 mW per pixel. The CCAT-prime RFSoC based readout is capable of reading four independent detector networks of up to 1000 KIDs each. The power dissipation was measured to be less than 40 W while running multi-tone combs on all four channels simultaneously. The system was also used for the first time to perform sweeps and resonator identification on a prototype 280 GHz array.
\end{abstract}
\blfootnote{
A.K. Sinclair, J.R. Burgoyne, S.C. Chapman, and R. Xie are with the University of British Columbia, Vancouver, BC  Canada (e-mail:adriansinclair@phas.ubc.ca). A.I Huber is with the University of Victoria, Victoria, BC Canada. Y. Li and C. Duell are with Cornell University, Ithaca, NY USA.
}

\section{Introduction}
Millimeter and submillimeter direct detector multiplexing methods experienced rapid growth and experimentation in the early 2000's \cite{Dobbs_2009}. Some of these early efforts have directly led to a substantial increase in detector counts for existing telescopes and some promise to enable even more for upcoming telescopes \cite{Parshley2018, SO2019, atlast2019}.

The Kinetic Inductance Detector (KID) was developed during this period  and employs a natural frequency division multiplexing \cite{Day2006}. Each KID is a superconducting thin film shunted resonator in which the absorption of photons above the band gap energy changes the inductance and thus the resonant frequency. These changes in frequency can be monitored as changes in transmission at the KIDs original resonant frequency. By designing an initial frequency spacing for each KID and coupling each to the same microwave feed line, frequency multiplexing factors above 1000 can be achieved with a single pair of coaxial cables. In addition to the high multiplexing factor the fabrication and cryogenic complexity is significantly reduced compared to other methods at the cost of higher complexity for the room temperature readout electronics.

KIDs were chosen as the ideal detector technology to efficiently fill in the wide field-of-view and large optical throughput of the upcoming Prime-Cam instrument on the 6 meter Fred Young Submillimeter Telescope (FYST) \cite{Vavagiakis2018}, previously known as CCAT-prime. Prime-Cam will observe the sky with over 100,000 mm/sub-mm direct detectors at 5600 meters in the Atacama desert of northern Chile. The nearly factor of ten increase in detector count from existing telescopes that Prime-Cam requires has spurred the development of a novel frequency multiplexing readout system. 

We present the first power dissipation measurements of the system and discuss the advantage of low resonant frequency KIDs. We also compare it to a previous generation readout and comment on the implications for planned and future balloon-borne mm/submm telescopes.

\section{RFSoC Readout and the Low Frequency Advantage}

The Radio Frequency System on Chip (RFSoC) developed by Xilinx has high speed digitizers, reconfigurable logic, and ARM microprocessors integrated all into a single chip. The integrated architecture has lowered the total power dissipation substantially when compared to equivalent implementations separated into multiple chips. All metrics of size, weight, power, cost, and bandwidth make it an ideal platform for the readout of frequency multiplexed detectors. An RFSoC based readout has been recently developed for Prime-Cam\cite{Sinclair2022}. 

All Prime-Cam KID arrays are being developed to resonate below the Nyquist frequency of the RFSoC digitizers allowing for direct RF sampling and generation. The digital design uses interpolation and decimation filters to limit the instantaneous bandwidth to 512 MHz and will use the numerically controlled local oscillator (NCLO) and digital mixers to up and down convert this band anywhere below Nyquist (2.048 GHz). This eliminates the earlier generation electronics requirement for analog mixers simplifying the warm readout considerably. The image tone suppression on the digital mixers were measured to be greater than 50 dB, a nearly impossible spec to meet with analog mixers. 

An additional benefit when choosing lower frequency resonators ($<$ 2 GHz) is that the number that can be multiplexed by the readout electronics increases with decreasing resonant frequency. This is due to the fact that the resonator bandwidth decreases with frequency $\Delta f = f/Q_r$. If the combined detector and readout electronics are viewed as an information carrying system then the low frequency advantage brings it closer to the Shannon channel capacity, although there is still a long way to go (see \cite{Irwin2009}).

Power dissipation within the RFSoC can be described by a static and dynamic term which scales with operational frequency. Increasing the instantaneous bandwidth of the readout requires either operating at a higher frequency within the fabric or  with an increased resource utilization. Both lead to increased power dissipation and thus provide another line of support for the low frequency advantage.

Some of the benefits described above were also found nearly a decade ago with detector performance advantages best described in Swenson et al. 2012 \cite{Swenson2012}.

\section{Power Dissipation Measurements}

\begin{figure}[!t]
\centering
\includegraphics[width=3.4in]{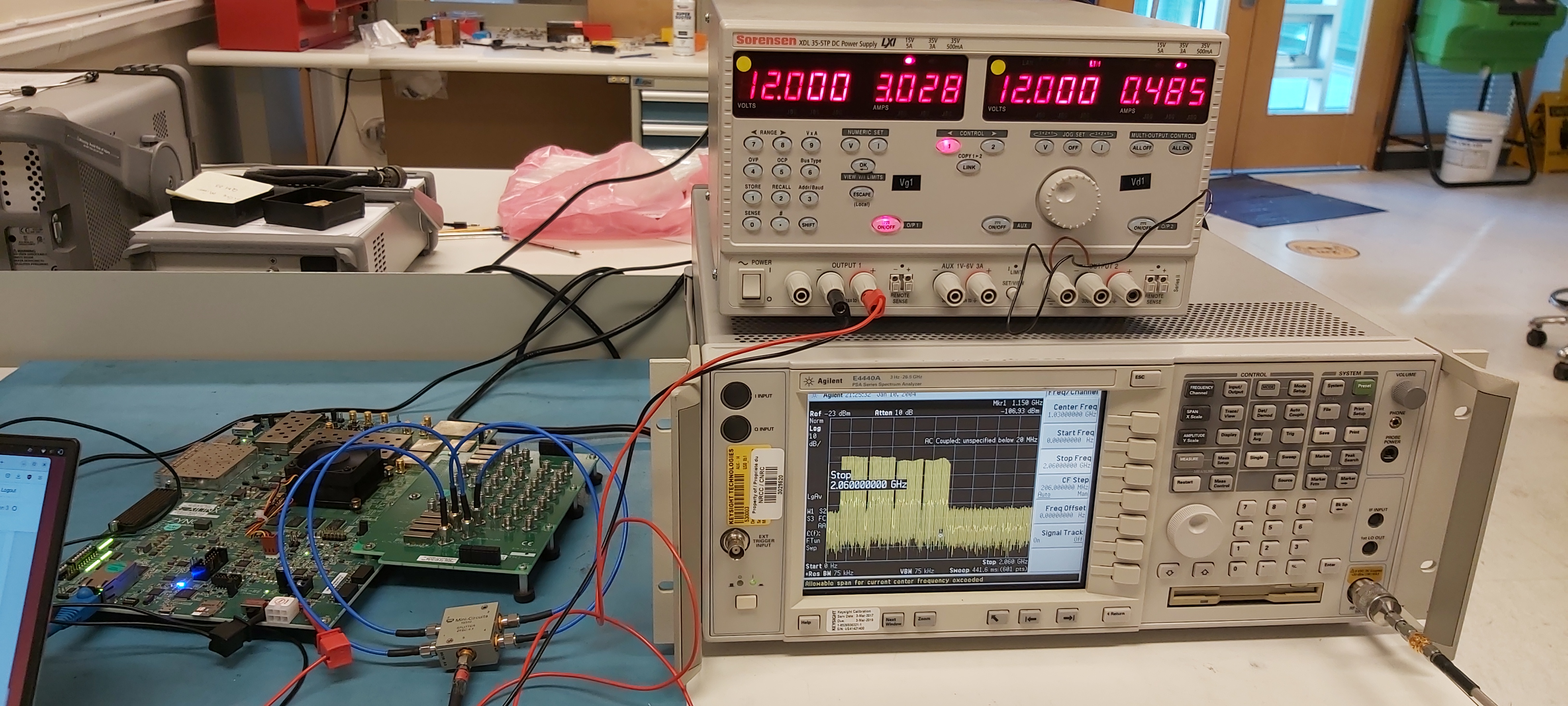}
\caption{Power dissipation measurement of four channel design. The RFSoC ZCU111 on the left was powered by the supply in the top right with 12 V and was drawing 3.028 A (36.34 W). The RFSoC was programmed with the four channel firmware and configured to generate frequency combs from each channel. Each frequency comb was digitally mixed with a different numerically controlled local oscillator frequency to separate the combs for visual display on the
spectrum analyzer.}
\label{fig:power_meas}
\end{figure}

Two power measurements were performed to verify the dissipation, one with a handheld multi-meter and the other with a benchtop power supply. 

First the ZCU111 development board power adapter output was measured with a multi-meter to be 12.23 V. The current was measured with a multi-meter in series between the output of the power adapter and the ZCU111 power Molex connector J52. Upon boot the system drew 1.83 A. In the boot state the ARM core loads a lightweight linux operating system and allows communication via Ethernet. The next step was to load the firmware configuration bitstream using the open source PYNQ library. After the bitstream was loaded the current rose to 2.7 A and stabilized. A frequency comb containing 1000 tones spanning 512 MHz of bandwidth were written to all four independent channels. The frequency comb waveform is normalized to maximize the available dynamic range of the D/A. With continuous operation of all four channels the current rose to 3.09 A and
stabilized. The total power draw for the multi-meter measurement was 37.8 W. 

The second power measurement was performed at the Herzberg Astronomy and Astrophysics Research Centre mm-wave lab with a DC power supply. The system was powered by applying 12 V directly to the Molex connector J52. After the bitstream was loaded and each channel was configured to generate multi-tone combs the current rose to 3.028 A and stabilized. This corresponds to a total power dissipation of 36.34 W. To show all four combs on a spectrum analyzer simultaneously a 4-way power combiner (Mini-Circuits ZFSC-4-1-S+) was found in the lab and connected to all four active D/A’s of the ZCU111 development board. Each numerically controlled local oscillator within the RFSoC were set to different center frequencies to spread the combs for visual effect. This measurement is summed up in one photograph show in figure \ref{fig:power_meas}.

\section{Comparison with Previous Generation Readout}
The author was involved in the development of an earlier generation KID readout \cite{Gordon2016} which has been deployed in two imaging polarimeter instruments at the Large Millimeter Telescope\cite{Rowe2022, Wilson2020}, and soon an on-chip spectrometer \cite{Redford2020}. It has also flown on two balloon-borne telescopes \cite{Olimpo2020, Coppi2020}. The digital signal processing algorithm used for these projects was the starting point for the RFSoC based readout system. 
	
The earlier generation readout mentioned above was implemented on the reconfigurable open architecture computing hardware version 2 (ROACH2) designed by the collaboration for astronomy and signal processing electronics research (CASPER) \cite{Hickish2016}. The gateware design implemented on the ROACH2 platform was capable of reading 1000 channels and drew 50 W (Not including intermediate frequency (IF) electronics such as the two analog mixers and external amplifiers), This corresponds to 50 W/1000 pixels = 50 mW/pixel. The new four channel RFSoC design can read out 4000 pixels for less than 40 W meaning $<$ 10 mW/pixel. BLAST-TNG flew five ROACH2s for a total of $\sim$ 250 W (not including IF) with a maximum possible number of detectors of 5000. This could be replaced with two RFSoCs running the four channel design for a significantly relaxed power dissipation, size, and weight. Utilizing all 250 W would correspond to six RFSoCs to read out a total of 24000 detectors. Future balloon-borne telescopes will benefit greatly from the RFSoC platform and a few are already planning to fly them \cite{Cataldo2021, Vieira2020}.

\section{First Measurement with Detector Array}

\begin{figure}[!t]
\centering
\includegraphics[width=3.4in]{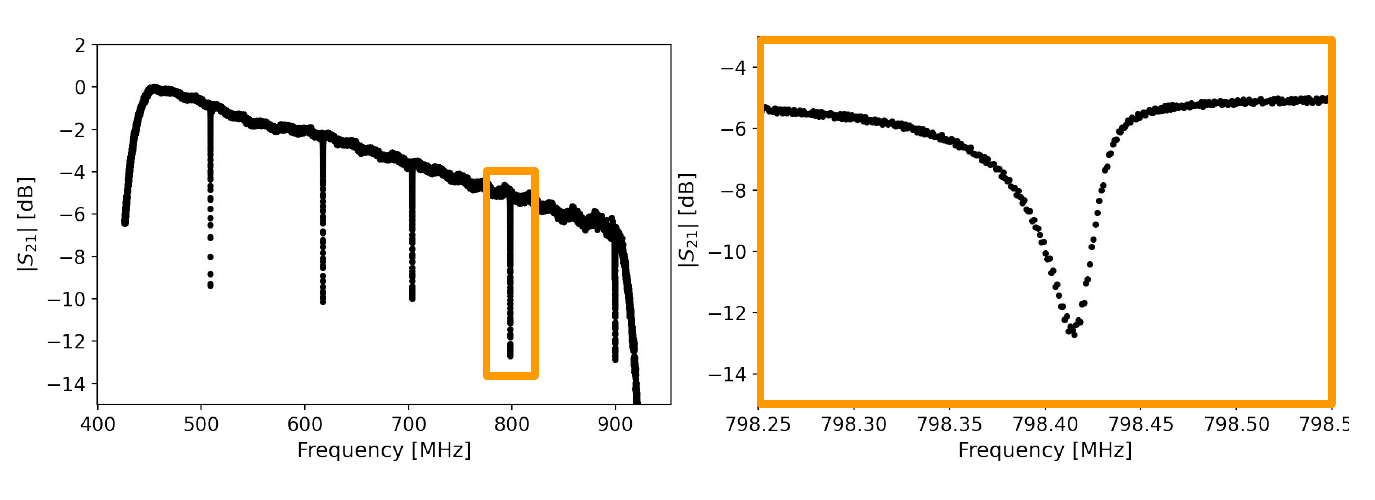}
\caption{Plots of the forward transmission as a function of frequency for a 280 GHz  prototype array measured by the RFSoC. The plot on the left shows the transmission normalized to the max for the operational bandwidth of 512 MHz and a numerically controlled local oscillator center frequency of 676 MHz. All five resonances corresponding to the five prototype detectors are easily identified. A zoom in of one resonance is shown on the right.}
\label{fig:detector_meas}
\end{figure}

Preliminary measurements were made with the RFSoC on a witness chip for the Aluminum 280 GHz array fabricated at NIST by the quantum sensors group. The array was cooled in a dilution fridge to 100 mK at Cornell University. The system contained 23 dB of cold attenuation to the devices and was followed by a low noise amplifier with 30 dB of gain and input referred noise temperature of 5 K. The RFSoC system was connected via an SMA breakout board to the cryostat with coaxial cables and a room temperature 10 dB fixed attenuator on the drive side. A 1000 tone comb was produced initially and measured on the spectrum analyzer to have approximately –50 dBm per tone. A preliminary version of the readout control software \cite{CCATpHive} was used to command the channel to generate the comb, perform a frequency sweep, and output a file containing the sweep measurement. The sweep used all 1000 tones and stepped the NCLO such that bandwidth between each tone was covered with a 1 KHz resolution. The result of the sweep is shown in figure \ref{fig:detector_meas} where each of the five detector resonances are identified and a zoom in of one resonator is also shown.

\section{Conclusion}
We have demonstrated a kinetic inductance detector readout that can achieve less than 10 mW/pixel. This system developed for the Prime-Cam instrument on the Fred Young Submillimeter telescope is ideal for KID-based balloon-borne telescopes. The systems on-board digital mixer and numerically controlled oscillator make transmission and resonator identification measurements with ease and considerably lower systematics. Future improvements in the readout gateware are anticipated and we expect the power per pixel to decrease even further on the same hardware. 

\section{Acknowledgment}
 The author would like to thank the mm-lab at the Herzberg Astronomy and Astrophysics Research Centre for the use of their space and measurement equipment. The CCAT-prime project, FYST and Prime-Cam instrument have been supported by generous contributions from the Fred M. Young, Jr. Charitable Trust, Cornell University, and the Canada Foundation for Innovation and the Provinces of Ontario, Alberta, and British Columbia.

\bibliographystyle{IEEEtran}
\bibliography{main}

\end{document}